\documentclass{article}
\usepackage{graphicx}
\usepackage{epstopdf}
\usepackage[all]{xy}
\newtheorem{theorem}{Theorem}[section]
\newtheorem{lemma}[theorem]{Lemma}

\newcommand{\qed}{\nobreak \ifvmode \relax \else
      \ifdim\lastskip<1.5em \hskip-\lastskip
      \hskip1.5em plus0em minus0.5em \fi \nobreak
      \vrule height0.75em width0.5em depth0.25em\fi}

\begin{document}

\title{How many independent bets are there in an emerging market?}

\author{Daniel Polakow\dag and Tim Gebbie\ddag \footnote{Department of Statistical
Sciences,  University of Cape Town, Rondebosch, 7700, South
Africa, and Peregrine Securities, PO Box 44586, Claremont, 7735,
South Africa, {\tt danielp@peregrine.co.za} \ddag Department of
Mathematics and Applied Mathematics, University of Cape Town,
Rondebosch, 7700, South Africa, \dag {\tt
tim.gebbie@physics.org}}}

\maketitle

\begin{abstract}
The benefits of portfolio diversification is a central tenet
implicit to modern financial theory and practice. Linked to
diversification is the notion of breadth. Breadth is correctly
thought of as the number of independent bets available to an
investor. Conventionally applications using breadth frequently
assume only the number of separate bets. There may be a large
discrepancy between these two interpretations.  We utilize a
simple singular-value decomposition (SVD) and the Keiser-Gutman
stopping criterion to select the integer-valued effective
dimensionality of the correlation matrix of returns. In an
emerging market such as South African we document an estimated
breadth that is considerably lower than anticipated. This lack
of diversification may be because of market concentration,
exposure to the global commodity cycle and local currency
volatility. We discuss some practical extensions to a more
statistically correct interpretation of market breadth, and its
theoretical implications for both global and domestic investors.
\end{abstract}

\begin{center}{\footnotesize{Keywords: Effective dimensions; Covariance Estimation; Emerging
Markets}} \end{center}

\section{Introduction}
One of the most widely accepted tenets of financial theory is the
principle that diversification is an essential component of any
well-constructed portfolio. Diversification serves to mitigate
specific sources or risk within any single asset class, and
systemic sources of risk across asset classes. Hence holding long
positions in two resource companies suchs as BHP Billiton and Rio
Tinto, may go a good way towards lessening the impact of
company-specific risk within the international resources sector.

Similarly, being exposed to property within a balanced (mutual)
portfolio lessens the threat that other asset classes
under-perform if property rallies. The idea is that spreading
one's bets results in value being unlocked slowly over time and
that diversification is a way to deal with an uncertain and
volatile investment universe. These are fairly convincing
argument's to most.

On a simple mathematical level, through diversification, one
enhances one's risk-adjusted return by nature of a principal
impact on any `risk' denominator, be it the standard deviation of
a Sharpe ratio or the active risk of an information ratio
\cite{Grinold1989,D90}. A portfolio that is comfortably
'diversified' is expected to have a higher Sharpe ratio and
information ratio. Diversification is frequently lauded as the
only free lunch that econometrics offers to fund managers, and
``one ought to indulge heartily at the price'' \cite{Thomas2005}.

It is from this optimistic base that we enter the fray with the
allegation that `diversification' opportunities may be both
limited and overstated.  Diversification in its common
pretext acts more to disguise value-add than to enhance it.
Interestingly, the usefulness of diversification in the way it was
originally intended is particularly limited in South Africa and
possibly other emerging markets for some less-than obvious
reasons, as we discuss later. As we illustrate, because
diversification is a frequently misunderstood phenomenon it can
clearly be a very mixed blessing.

To the skilled fund-manager diversification may actually be an
impediment. Spreading one's bets too thinly across independent
gambits condemns such talent towards the manifestation of
mediocrity since there is little room to move efficiently in all
dimensions. To the less prudent fund manager, however,
diversification will often offer a safe-haven where poor bets
amongst will be simultaneously countered by good ones in others.
Furthermore, we show that in the context of conventional asset
classes within Southern Africa, there is a lot less room to
maneuver than most professional investors suspect due to an
overriding communality of extraneous factors that impact similarly
on a wide variety of asset classes.

This has both implications for those global investors naively
treating emerging markets as an independent asset class, those
seeking international diversification from within an emerging
market, and attempts to understand the theoretical applicability
of asset pricing models in emerging markets.

In section \ref{ss:breadth} we discuss `breadth'; how it is
understood, used, and typically misused. This is followed in
section \ref{ss:svd} by a discussion of well-known multivariate
statistical technique that facilitates a more correct
understanding of the available breadth within any universe of
assets: using the singular value decomposition in conjunction with
the Keiser-Gutman stopping criterion, to select the integer-valued
effective dimensionality of a correlation matrix of returns, with
eigenvalues greater-than or equal to 1. We are then able, in
section \ref{ss:examples} to illustrate the available breadth to
investors in South African markets, by using some well known
multivariate statistical techniques in relation to three examples:
a portfolio of equity (See Figure's \ref{fig:figure1} and
\ref{fig:figure2}), an equity and bond portfolio (See Figure's
\ref{fig:figure3} and \ref{fig:figure4}) and last, a portfolio
including, in addition to equity and bonds, cash,
property and international bonds and international equity (See
Figure \ref{fig:figure5} and \ref{fig:figure6}).

Lastly, in light of the insights provided from these previous
chapters, in section \ref{ss:conclusion} we reconsider the role
that asset allocation has within the context of a resident
balanced portfolio and also focus our discussion within the
context of the useful fundamental law of active management
\cite{Clarke2002, Grinold1989}.

\subsection{Breadth - Independence rather then separateness?}
\label{ss:breadth}

Conventional theory suggests that an increase in diversification
opportunities (N) is accompanied by an increase in one$'$s
information ratio (IR) \cite{Grinold1989}. Hence, in the
terminology of active management, an increase in N serves to
enhance ones ability to exploit information.  Note at the onset
that N is defined and treated as the number of separate bets
(sensu Clarke et al. 2002).

For example, assume we have a 60\% chance of getting equity bets
correct. A bet on one underlying will yield an IR of 0.2, a bet on
five underlying securities, an IR of 0.45 and a bet on 20
underlying securities, an IR of 0.90. This situation is easily
verified. The understanding stemming from this detail is
universal. For example, Lee Thomas \cite{Thomas2005} notes the
following implications for considering diversification in the
selfsame light:
\begin{enumerate}
\item Since diversification has an obvious statistical basis, a
larger number of bets will produce a higher information ratio.
\item A lot of the differences between fund managers' performances
are often ascribed to `skill' whereas the differences may simply
be an artifact of better diversified portfolios.
\item The search
for higher quality investments should be superseded by the search
for diversified investments.
\item Diversification is paramount to
investment success – across asset classes, styles and countries.
Diversify, diversify, diversify!
\end{enumerate}

The above-mentioned argument's are very appealing, and very well
utilized, but we disagree with each and every contention since all
omit two essential truisms, which when understood, shed a very
different light on the benefits of diversification and the nature
of breadth.

\begin{lemma}[Independence is not separateness] \label{lemma:truism1}
The square root of N in mathematical statistics
implies `independence' amongst statistical units (here bets)
\cite{Rice1995} rather than simply the notion of `separate bets'
as is most often implied.
\end{lemma}

 If I hold a portfolio of 10 single
stocks, do I really have 10 `independent' bets and is my breadth
really 3.16? If I increase this to 1000 single stocks (assuming I
have as many available), is my breadth 31.6?  This is the key
theoretical question dealt with here.

\begin{lemma}[Skill does not scale over breadth]\label{lemma:truism2}
Skill is not generally or simply scalable over breadth. One
requires considerable skill in preserving one's information
coefficient (or IC) \cite{Clarke2002} across an increasingly
diverse universe of investable underlying securities.
\end{lemma}
There is no {\it a priori} reason to expect, for
argument's sake, a South African fund manager or analyst to be as
adept in understanding the earnings potential of a diversified
industrial company, as in understanding the risks and upside of
Chinese private equity; yet there is a continuity of forecasting skill
invoked across both. This is a key practical implication considered here.

Taken to its logical extreme, there is no reason to expect the
same information coefficients (IC) between any two underlying
securities. IC is an average measure that is typically applied to
the sum total of all bets in a portfolio. We need to disaggregate
the measure to understand its scalability.

Understanding the reality and the benefits of diversification
resides in understanding both truisms given in Lemma (\ref{lemma:truism1})
and Lemma (\ref{lemma:truism2}) concurrently. We focus on the following
three pertinent questions:

\begin{description}
\item [Question \#1] Just how many South African single stocks do I need to add to
a portfolio before I start to replicate pre-existing elements of
diversification (i.e. saturate all elements of independence)?
\item [Question \#2] Do I capture much more `breadth' if I include other local and
international asset classes?
\item [Question \#3] What are the implications of
these findings to fund managers?
\end{description}

\subsection{Methodology - The informational content in a SVD}
\label{ss:svd}

The foundation of this research arises from the confusion between
the notions of `separate bets' and `independent bets'; the two
are not the same. The question really is - how alike are they, and
are there better ways for understanding independence than
through the manner in which \cite{Grinold1989,Clarke2002} imply?
We believe there are several ways to better represent independence
than through Grinold's original construct.

For our purposes, as well as for ease of replication, we make use
of the principle of `effective dimensionality': given a return
matrix $X$, we us the singular-value decomposition to factorize
$X$ as $X = U \Sigma V^{_T}$ for eigenvectors $\Sigma =
\mathrm{diag}(\Sigma)$ and eigenvalues $U$. The unitary matrix $U$
spans the subspace where the variations in the data are the
largest. Each eigenvalue has an associated eigenvector. We then
utilize the Keiser-Gutman \cite{Jackson1993} stopping criterion to
select those eigenvectors with eigenvalues greater than or equal
to one, the number of such eigenvalues corresponds to an
estimation of the effective dimensions of the subspace - the N in
the fundamental equation of active management
\cite{Grinold1989,B90}.

There are alternative means of defining the effective number of
stocks in a portfolio using entropy measures \cite{Fernholz1999}.
These all ultimately revolve around the degree of localization of
the portfolio controls under optimization, and hence pivot on the
appropriate definition independence as differentiated for
separateness. Many definitions of entropy assume, {\it a priori}
independence amongst the states (here bets) of the system, this as
we argue may in fact end up being problematic in these new
reformulations of the notion of effective number of bets if not dealt
with sensibly.

\subsection{Empirical example of the enhanced application}
\label{ss:examples}

We utilize return data from the Johannesburg Stock Exchange (JSE)
and the Bond Exchange of South Africa (BESA) for the purposes of
demonstrating both our breadth computations as well as the
evidenced effect of limited breadth within the South African
marketplace.  We use daily data for a period of $4.3$ years (March
2003 - present) for $41$ of the most liquid equity stocks on the
JSE. The most liquid equity index is termed the Top-40 index. The
period of $4.3$ years is arbitrarily chosen as a cut off point
where most of the equity counters currently trading are subsumed
in the analysis.

\begin{figure}[h]
\includegraphics[width=11cm]{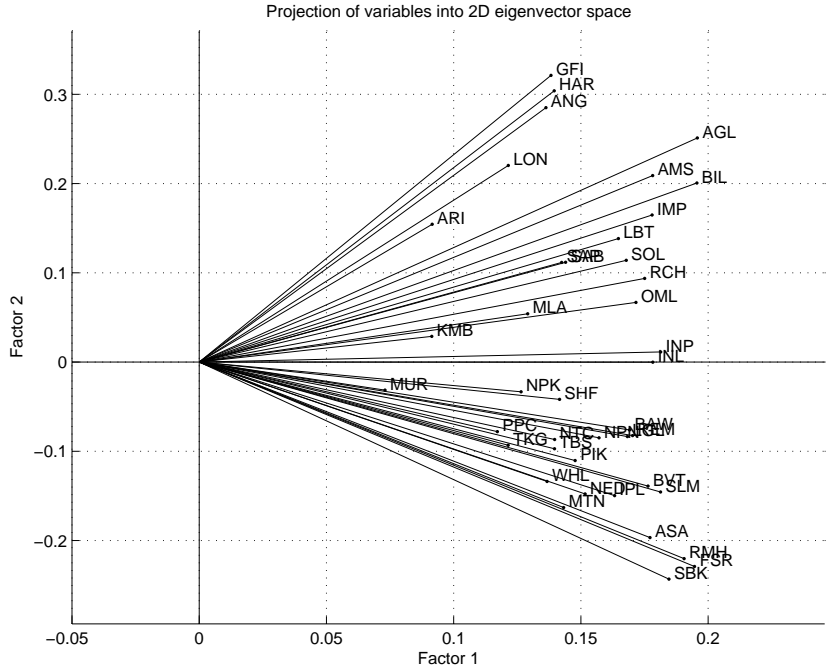}
\caption{Example \#1: The projection of the single-stocks
(variables) onto the 2-D eigenvector space is provided as a
bi-plot for. Note the prevalence of the bulk of the counters in two of
the four quadrants.  Note also that banking and gold-mining stocks
cluster at the two end extremes.} \label{fig:figure1}
\end{figure}

We commence by computing the effective dimensionality of this
sample of $41$ single-stocks from the estimated correlation
matrix. A projection of the single-stocks (variables) onto the 2-D
eigenvector space is represented in Figure \ref{fig:figure1}. The
projection shows that some gold stocks (ANG, GFI and HAR) cluster
together at the extremes of the first two eigenvectors. Similarly,
most banking stocks (e.g. SBK, FSR, ASA and RMB) cluster at the
opposite quadrant of the same two eigenvectors.

\begin{figure}[h]
\includegraphics[width=11cm]{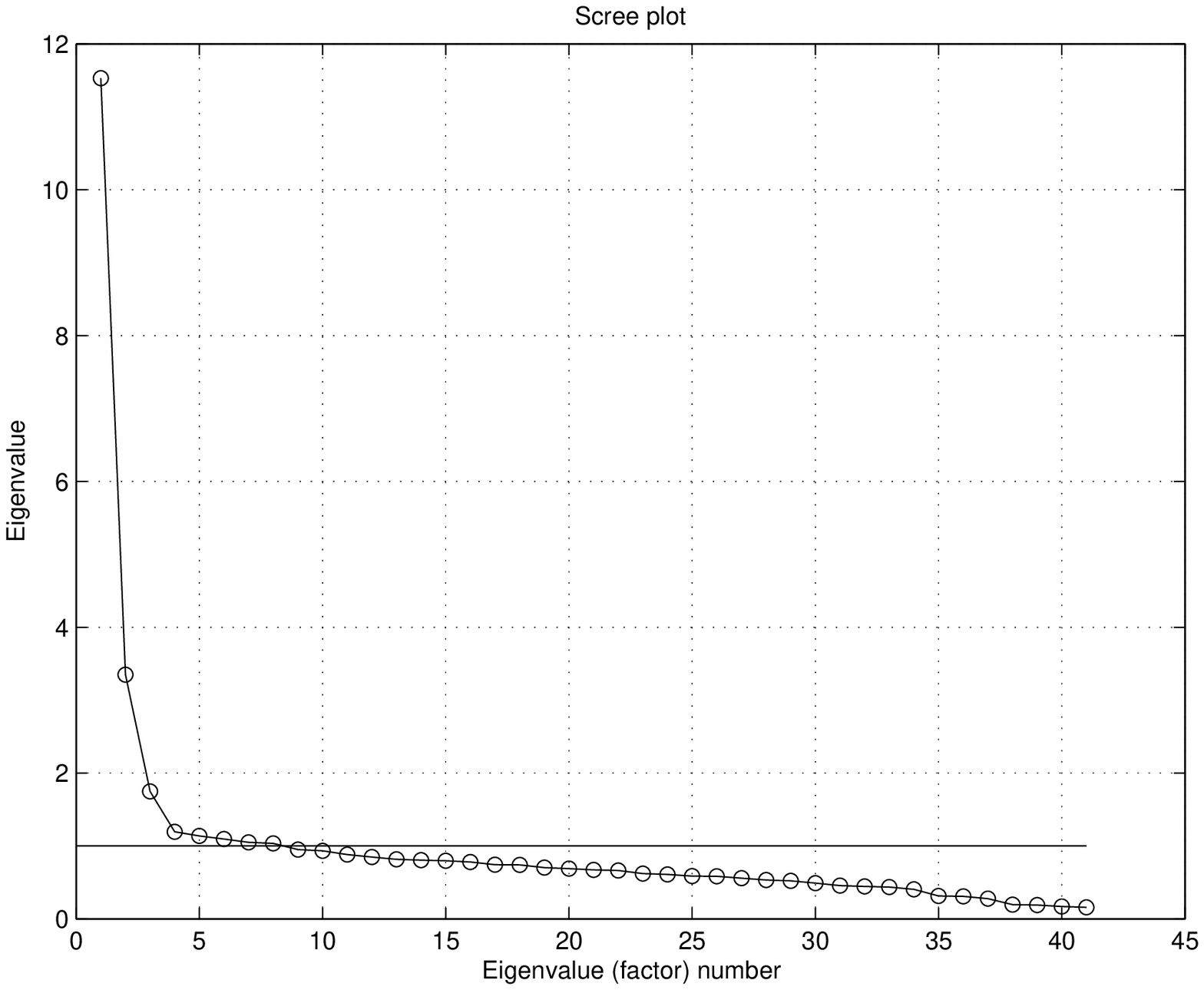}
\caption{Example \#1: The scree-plot mapping of the decay of the
eigenvalues by the dimensionality of the single equity stock universe.
The effective dimensionality of the data-set is found to be no more
than $8$ dimensions when using the Keiser-Gutman criterion. This
is an effective breadth of approximately $3$.} \label{fig:figure2}
\end{figure}

A scree-plot is used to map the decay of the eigenvalues by the
dimensionality of the data set in Figure \ref{fig:figure2}.

Using our Keiser-Gutman criterion, we compute the effective
dimensionality of the dataset as no more than $8$ dimensions. This
translated into an effective breadth of about $3$.  The
conventional use of the fundamental law of active management would
estimate breadth here at $\sqrt{41} = 6$, twice that evidenced
here.

\begin{figure}[h]
\includegraphics[width=11cm]{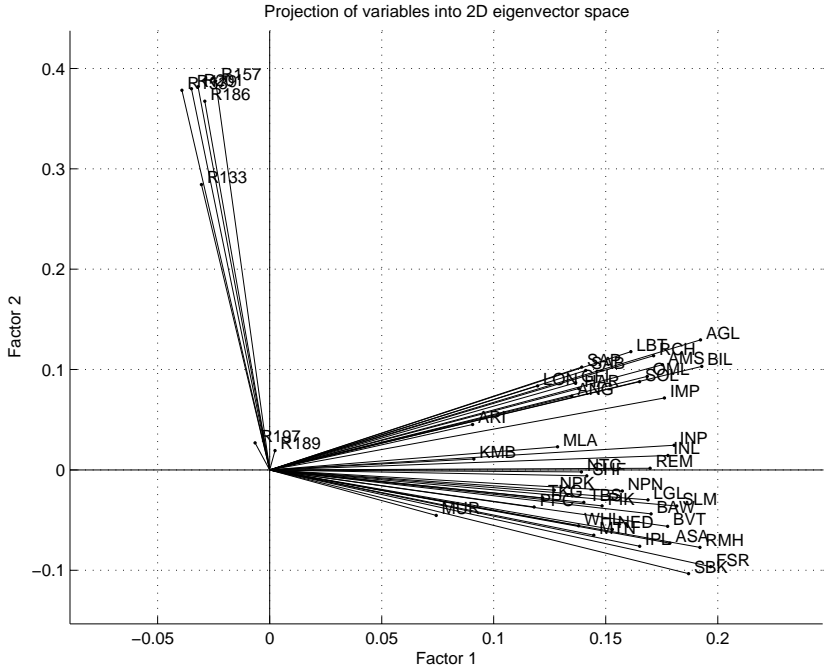}
\caption{Example \#2: A projection of the underlyings onto the 2-D
eigenvector space. We include 8 dominant
government-issued bonds along with the 41
single-stocks.}\label{fig:figure3}
\end{figure}

Next, we repeat the selfsame exercise as above, but now consider
jointly the selfsame period of eight total return series of the
seven dominant government-issued bonds along with our 41
single-stocks. A projection of the underlying securities onto the
2D eigenvector space is noted in Figure \ref{fig:figure3}.

\begin{figure}[h]
\includegraphics[width=11cm]{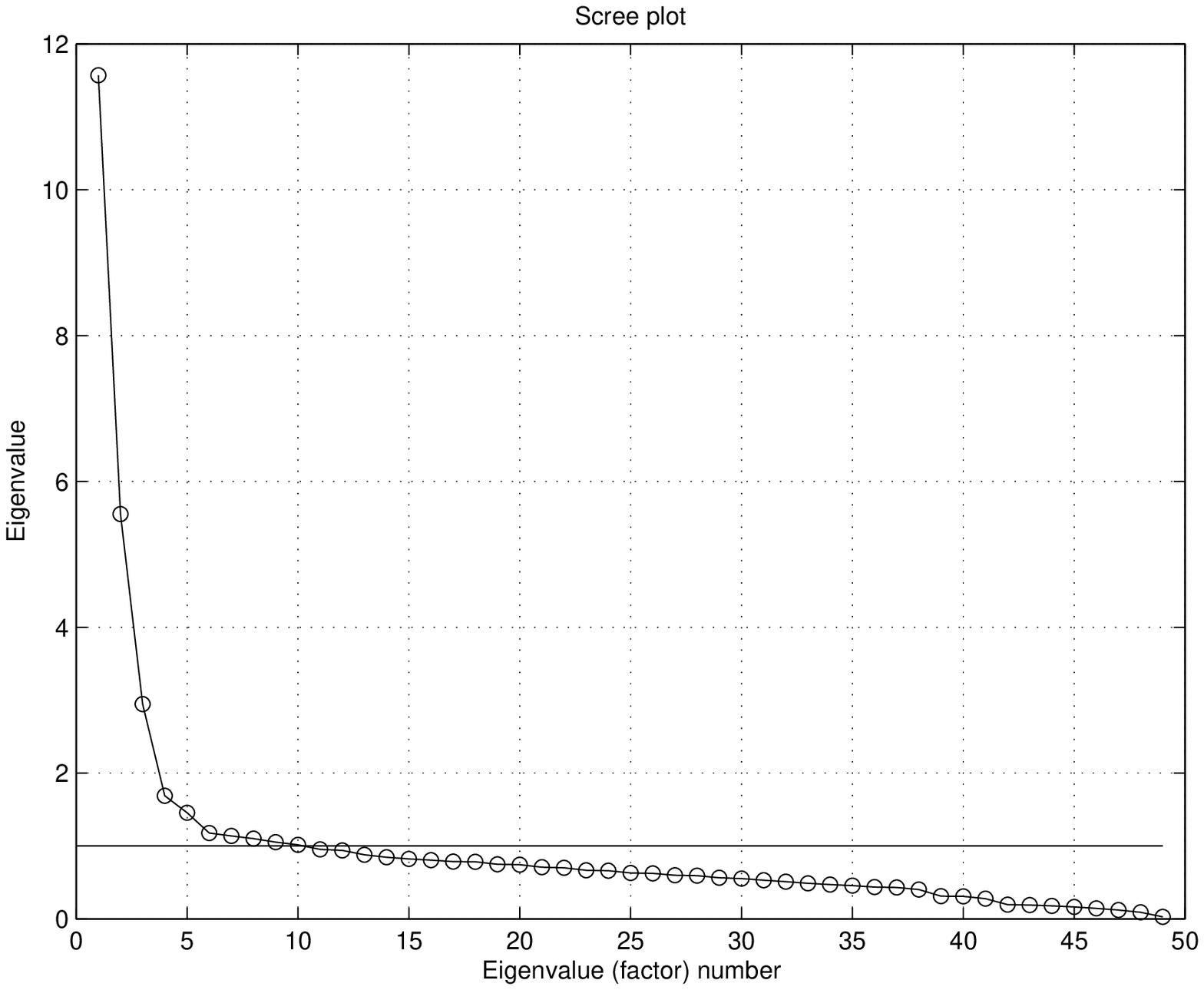}
\caption{Example \#2 : The scree-plot showing the decay of the
eigenvalues in the equity-bond universe. The effective
dimensionality is estimated as no more than 9 dimensions,
translating into an effective breadth of 3.}\label{fig:figure4}
\end{figure}

In Figure \ref{fig:figure4} a scree-plot is once again used to map
the decay of the eigenvalues by the dimensionality of the data
set.

The effective dimensionality is estimated as no more than 9
dimensions, translating into an effective breadth of 3.
Conventional analysis here would infer a breadth of 7.  Note how
the analysis suggests both that South African bonds do not present
much of a diversification enhancement to an equity portfolio and
that replication of the self-same communalities exist.  The
reasons for these anomalies are easily explained by the dominant
role that the local exchange rate plays on both equity and bond
valuation and the impact of interest rates and commodity pressures
on both.

\begin{figure}[h]
\includegraphics[width=11cm]{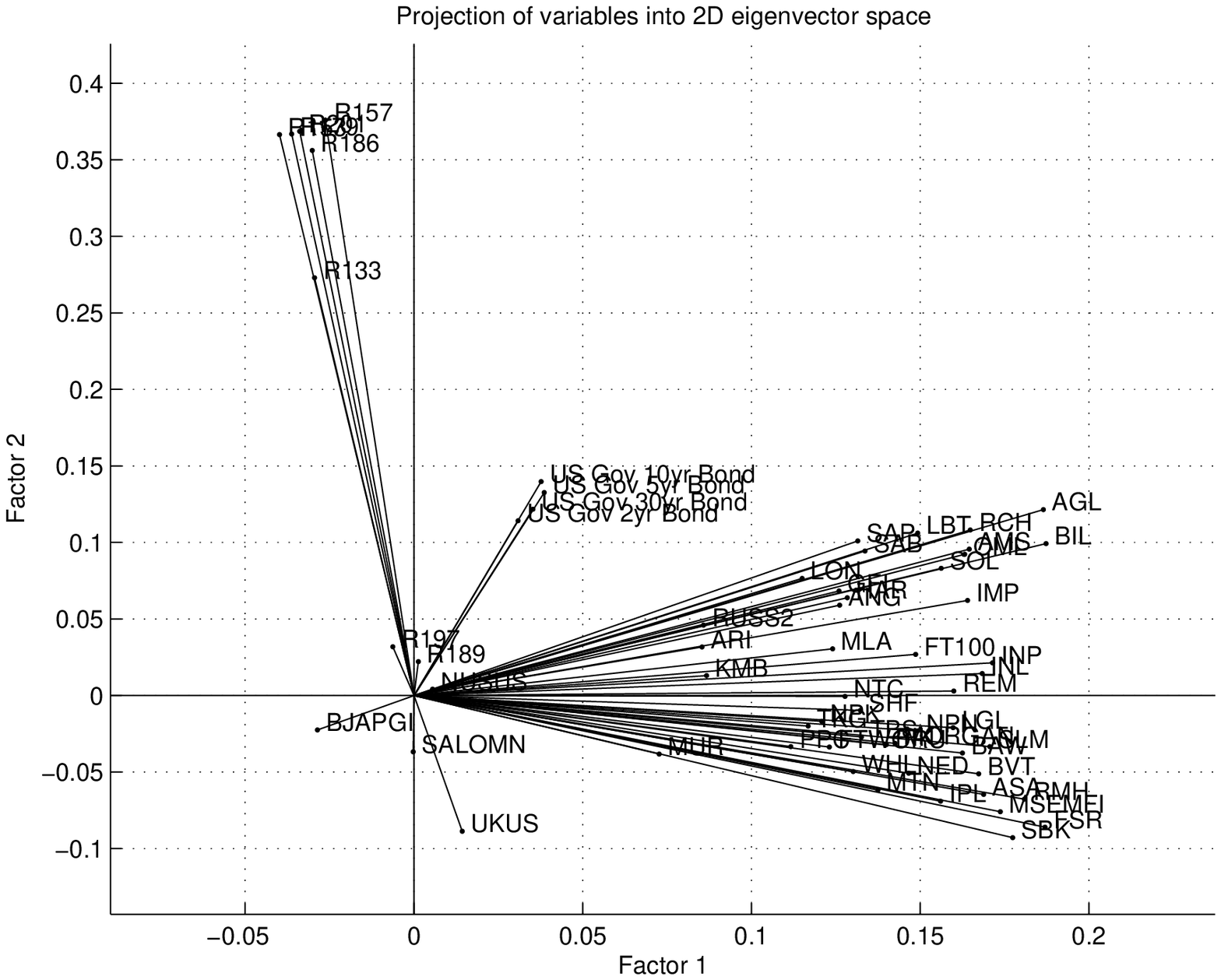}
\caption{Example \#3: A projection of the underlyings onto the 2-D
eigenvector space. We include 13
international bonds and equity indices along with the 41
single-stocks and 8 government South African issued
bonds.}\label{fig:figure5}
\end{figure}

Lastly, for the purposes of illustration, and to ascertain the
place that the South African marketplace assumes relative to
international markets, we include 13 new international assets to
our analysis, all hard-currency denominated, including both bond
indices and equity indices notably: the FT World Equity Index
(USD), the MSCI World Equity Index (USD), four US Government bonds
of varying duration (USD), Japanese inflation-linked bonds (JPY),
a US corporate bond index (USD), an Emerging Market Index (USD),
the FTSE 100 (GBP), the Russell 2000 (USD) , the Citigroup world
government bond index (USD) and the UK government bond index
(GBP). A projection of the underlyings onto the 2D eigenvector
space is noted in Figure \ref{fig:figure5}.

It is interesting to note from \ref{fig:figure5} that most of the
new (international) assets span separate dimensions to the ones
the South African securities occupy, apart for two obvious
exceptions. First, US bonds are related to South African bonds -
the correlations between these are strong and positive (as
indicated by the acute angle between the bi-plot radians).
Second, South African single-stock underlyings are inextricably
tied to international equity markets - as evidenced by the
presence of both the FTSE World and MSCI World Equity Indices in
the same quadrant and in the same direction, despite the
hard-currency differences here.

\begin{figure}[h]
\includegraphics[width=11cm]{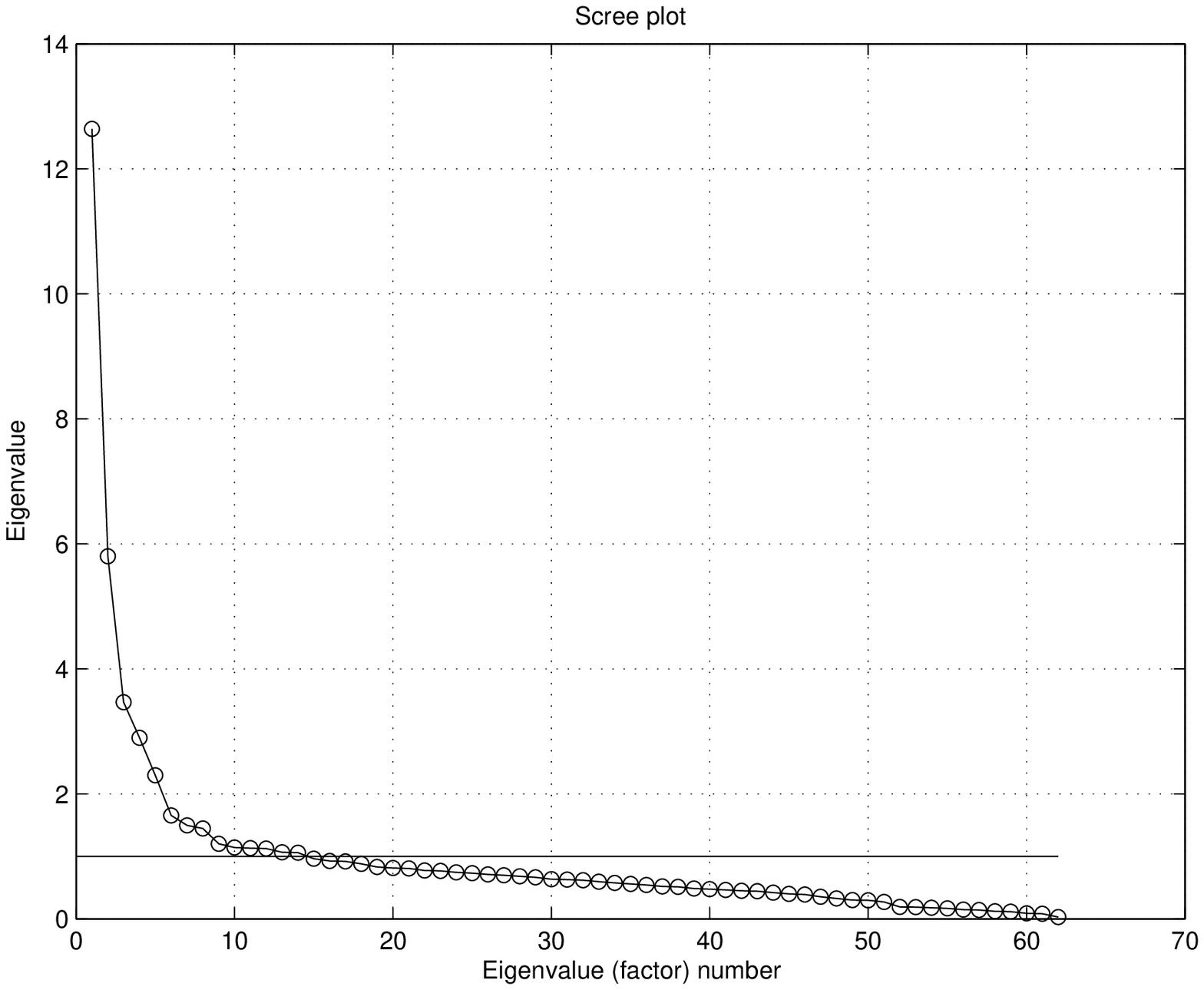}
\caption{Example \#3 : The scree-plot showing the decay of the
eigenvalues in the mixed universe. The effective dimensionality is
estimated as no more than 13 dimensions, translating into an
effective breadth of close to 4.}\label{fig:figure6}
\end{figure}

In Figure \ref{fig:figure6} a scree-plot is once again used to map
the decay of the eigenvalues by the dimensionality of the data
set. Including $13$ of what most investors would consider to be a
fairly diverse range of international asset classes to our
original dataset of $41$ South-African equity underlyings together
with eight South African government issued bonds increases the
effective dimensionality by only 4(from 9 to 13). Conventional
methods would impute a breadth of closer to 8 whereas the actual
breadth is closer to 4.

\section{Conclusion} \label{ss:conclusion}

Breadth in our approach takes on a meaning that is quite different
from that usually used, but is closer to the spirit of the idea, as
it measures the benefit of real diversification. The SVD approach
to the problem handles independence of the basis of spanning
eigenvectors correctly whereas the notion typically assumed, that
gambles are independent by their very nature, is simply incorrect.

One of the key results arising from this investigation into
diversification possibilities in an emerging market such as South
Africa is the significant limitation of breath in the opportunity
set. This breadth limitation may well be due to a concentration of
capital in a handful of local stocks within the South African
market. Some 33\% of the market capitalization is contained in the
top-5 underlying securities. Interestingly, most of these stocks
are dual-listed overseas, hence further limiting the breadth
expansion when international assets are included in a resident
South African portfolio. This breadth limitation has implications
for the variety and nature of the risk associated with investment
strategies and questions conventional wisdom in the construction
of global funds.

A common misconception prevalent in the literature regarding the
benefits of diversification is that skill is scaleable over
breadth \cite{Thomas2005}. Hence, diversification is a free lunch
offered to a `diversified' portfolio in the sense that a larger
number of bets effected with the same skill will produce higher
IRs. However, the error is made in assuming that one's IC remains
constant as breadth increases. Clearly, it cannot. For every added
dimension of independence, one seems to require a novel skill-set.
In the context of asset management, the implications of this error
in the context of a limited breadth debate are threefold.

First, real diversification into breadth to achieve an optimal
risk-adjusted return (via the IR) requires skill (as gauged by
IC). If IC is compromised by breadth increasing, as we expect it
to be, it can be argued that `diversification' is actually a
recipe for mediocrity amongst professional fund managers in the
most general case.  A prefatory analysis of several South African
fund managers show different levels and persistence of IC for
different sectoral bets.  It would be of specific interest to
quantify where the value resides within such institutions, and how
to best extricate this value-add in the context of a balanced
(mutual fund) mandate.

Second, less breadth will exist within any one asset class than
the fundamental law implies.  For a unit of capital, shifting
allocation within an asset class will increase the breadth less
(if at all) than shifting allocation across asset classes (the
idea of tactical asset allocation). In the context of the South
African marketplace, limited diversification exists within a
highly concentrated equity market such that a shift from one
security to another represents more of a bet about the relative
spreads between their expected returns than it does anything about
diversification. In fact, a pair-trading strategy (in its own
right) creates a dimension of independence that is uncorrelated
with either of the two original underlying securities, but may be
correlated with other positions.

Lastly, it should be noted that tactical asset allocation can
facilitate a rapid breadth expansion by translating 'possible'
breadth into `realized' breadth.  The pros and cons of asset
allocation need to be considered in the selfsame context of the
skill that managers have in timing the movements in various asset
classes versus the diversification benefits of so doing.  In this
sense, our proposed modification to the fundamental law of active
management provides a generalizable framework in which both
static, dynamic and tactical asset allocation can be thoroughly
and correctly investigated. In this context, IC(t), the
information coefficient as a function of term, is the basis on
which any analysis needs to be focussed.

The prospects of utilizing the fundamental law in this manner are
particularly piquant and we hope that this research will stimulate
some further work in this area. We note that the coefficients that
are ultimately derived form the fundamental law of active
management will not be comparable with previous studies, where
breadth has not been estimated correctly.

\section*{Acknowledgments}

The MATLAB code and the data used to produce the graphs can be
obtained from DP. The authors thank Mark De Araujo, Diane Wilcox
and Rayhaan Joosub and for helpful insights and suggestions, comments and
criticism.

\end{document}